# DYNAMIC EXTRACTION OF KEY PAPER FROM THE CLUSTER USING VARIANCE VALUES OF CITED LITERATURE

Akira Otsuki[1]

[1] Tokyo Institute of Technology, Tokyo, Japan

cecil2005@hotmail.co.jp

## ABSTRACT

*When looking into recent research trends in the field of academic landscape, citation network analysis is common and automated clustering of many academic papers has been achieved by making good use of various techniques. However, specifying the features of each area identified by automated clustering or dynamically extracted key papers in each research area has not yet been achieved. In this study, therefore, we propose a method for dynamically specifying the key papers in each area identified by clustering. We will investigate variance values of the publication year of the cited literature and calculate each cited paper's importance by applying the variance values to the PageRank algorithm.*

## 1. INTRODUCTION

We became able to analyze the academic landscape by the "Citation Maps" based on the bibliometrics or clustering etc. The citation map is the network map that shows citation relationship of the journal papers. And the nodes in the citation map are divided into groups by clustering method. We became able to grasp than before the recent research trends by citation map. But the reality is that experts manually analyze to identify the type of each cluster (group) that divided by clustering. Therefore, dynamically interpretation of each cluster is the final purpose of our study, and this study will achieve an automated extraction of key literature of each cluster. Concrete, we propose a method for dynamically specifying the key papers in each area identified by clustering. We will investigate variance values of the publication year of the cited literature and calculate each cited paper's importance by applying the variance values to the PageRank algorithm. Furthermore, we will attempt to create a visualized graph with the time axis based on the importance. This paper is composed of the following sections: Section 2 provides an overview of prior research on element techniques for achieving academic landscape, such as network analysis, clustering, and bibliometrics; Section 3 describes the purpose of this study and the proposed method; Section 4 describes the results of the evaluation tests of our proposed method; and Section 5 offers a conclusion.

## 2. RELATED STUDIES

### 2.1. Network Analysis

The academic landscape discussed in this study is a type of network analysis. It is the conception to analyze visually the enormous journal papers by analyzing the citation network and clustering of papers published in academic journals. By giving in each cluster common features, it is possible to simplify the overall structure of complex data. Then it will be able to divide into each research areas as shown in Fig1.

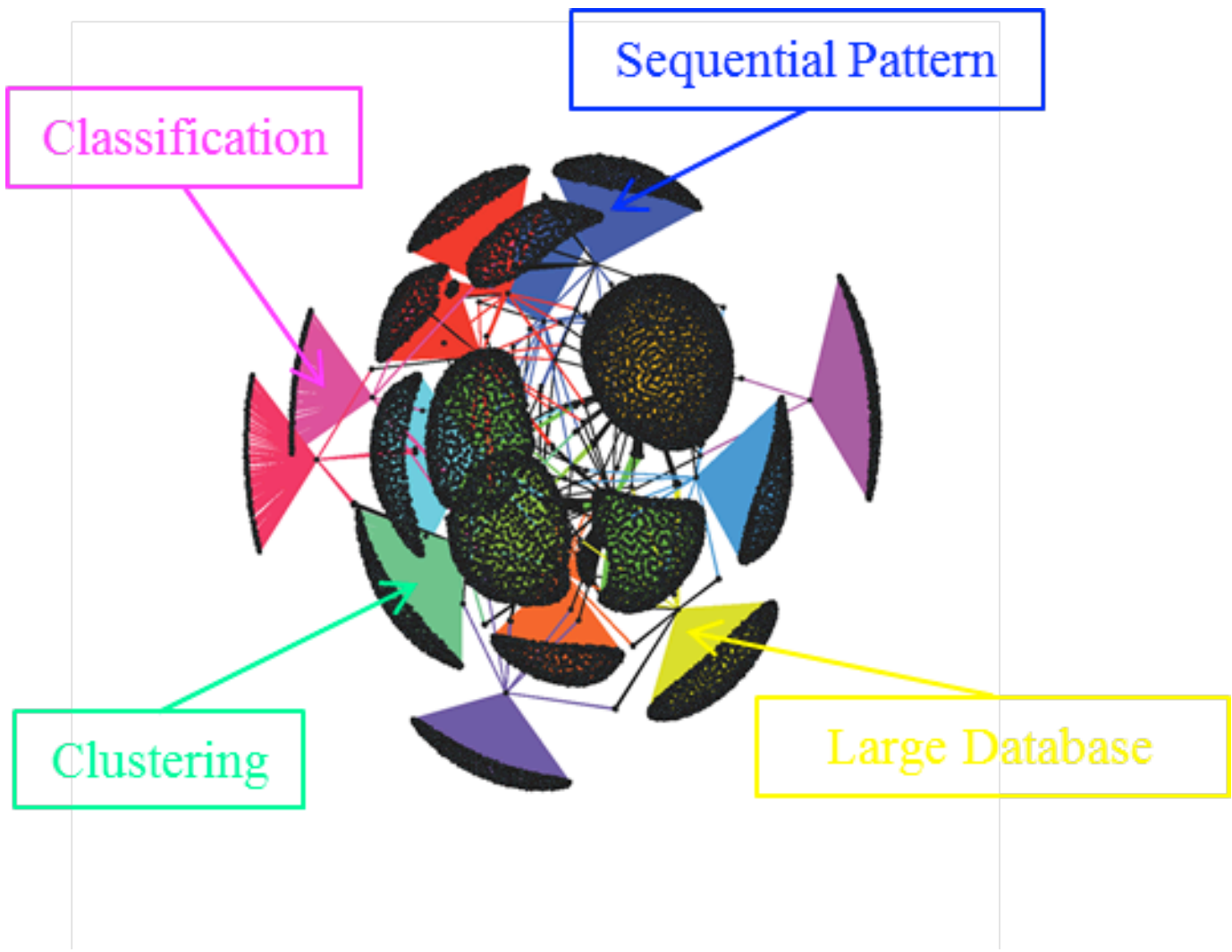

Figure 1. Example of academic landscape

While the study of network analysis is an area with a long history, dating back to graph theory proposed by L. Euler in the 18th century [1], this area is said to have achieved great advancement only in the last ten years. Network analysis involves various techniques including calculation of stele, contention analysis by structural equivalence [2], centrality [3-4], small-world property [5-6], degree distribution, and clustering [7-9]. Among these, the technique that is currently closely related to academic landscape is clustering. Clustering is described in the following section.

## 2.2. Clustering

The characteristics of clustering are based on graph theory. By giving the data contained in each cluster common features, it is possible to simplify the overall structure of complex data and understand the data more thoroughly (Fig1). In the initial clustering algorithm, the commonly used technique focuses on the "central" link and links are cut one by one starting from the central one. On the other hand, Girvan and Newman focused on the links that mediate the clusters nearest to other clusters using modularity as the evaluation function and proposed an algorithm to cut links in descending order of their mediating power.

## 2.3. Bibliometrics

Bibliometrics is a technique developed by Garfield [10-12]. He proposed the Science Citation Index in the 1950s as a tool for use by scientists to retrieve early science research. The Science Citation Index was developed into bibliometrics. Bibliometrics makes academic papers and patents the object of analysis and enables analysis on issues such as "what is a hot topic," "which paper is cited more frequently," "what field is related to what filed," and "who is an important researcher" in a certain research area by analyzing the papers using quantitative analysis techniques. Bibliometrics involves three analysis techniques as described below (Fig2). The first analysis techniques is direct citation; in Fig2, Papers A and B are cited in Paper C, and Paper C is cited in Papers D and E. In this case, direct citation deems that there are links between Papers A/B and Paper C and further links between Paper C and Papers D/E. As a result, there are five nodes and four links in the network. When direct citation is used, a certain paper is deemed to have links with all papers that cite the pertinent paper. The second technique is co-citation, which was proposed by Small [13]. In Fig2, both Paper A and Paper B are cited in Paper C. In this case, co-citation deems that there is a link between Paper A and Paper B; thus, there are two nodes and one link in the network. For pairs of papers in which co-citation was used, i.e., all papers contained in the list of cited literature of a certain paper, there is a link between the paired papers. The third technique is bibliographic coupling—a technique proposed by Kessler [14]. In Fig2, both Paper D and Paper E cite Paper C. In this case, this technique deems that there is a link between Paper D and Paper E; thus, there are two nodes and one link in the network. When bibliographic coupling is used for pairs of papers that cite a certain paper, it is deemed that there is a link between the paired papers.

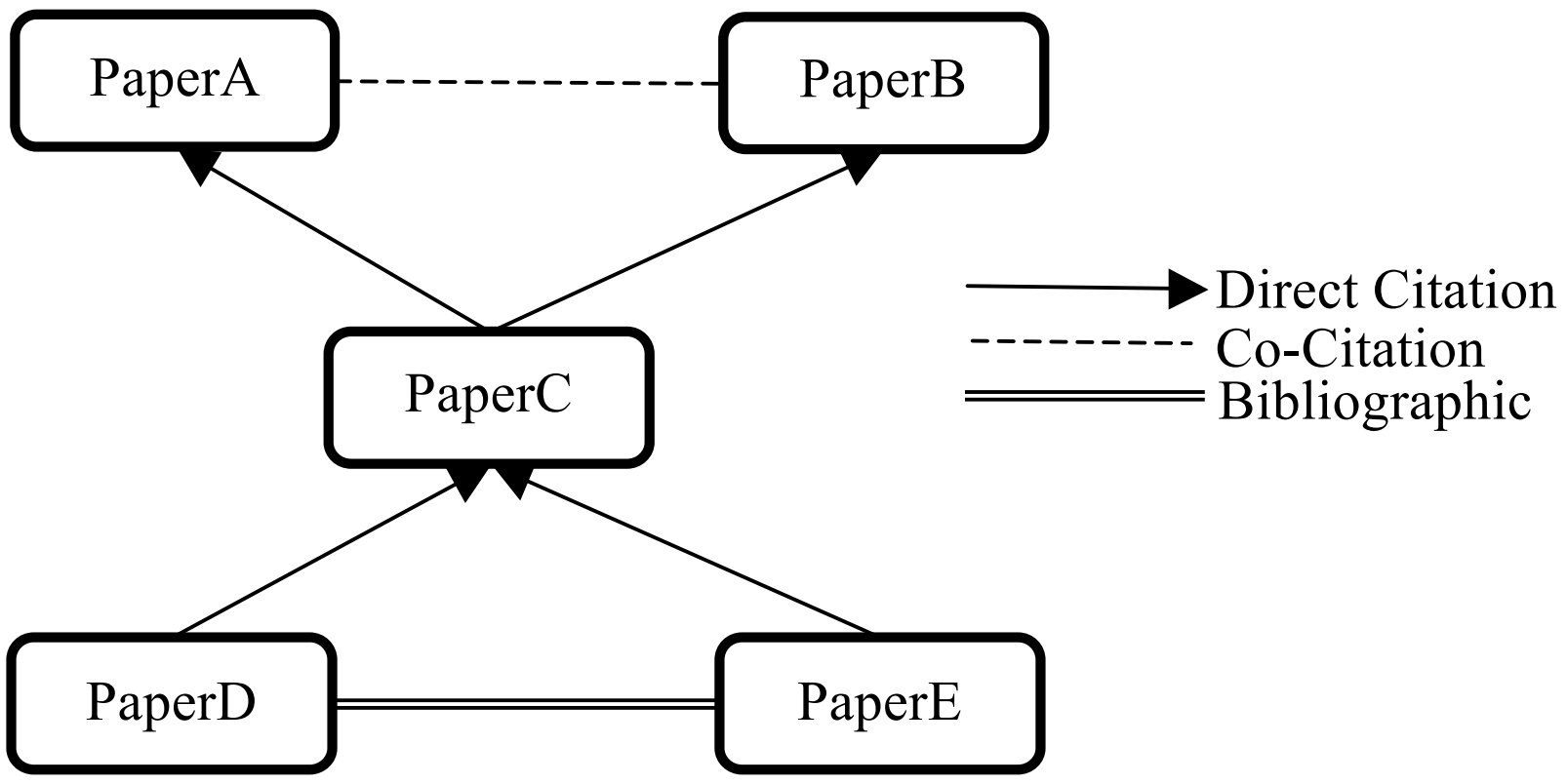

Figure 2. Three analysis techniques in bibliometrics

## 2.4. The Issues of Academic Landscape

Through a combination of the various techniques described in the previous section, it is possible to analyze the academic paper citation network by construct of the citation networks, acquisition of the largest connected component, clustering, and visualization dynamically, as shown in Fig3. However, automation is not yet implemented for specification of the areas identified by clustering or extraction of key papers and key researchers and this portion is currently manually analyzed by experts.

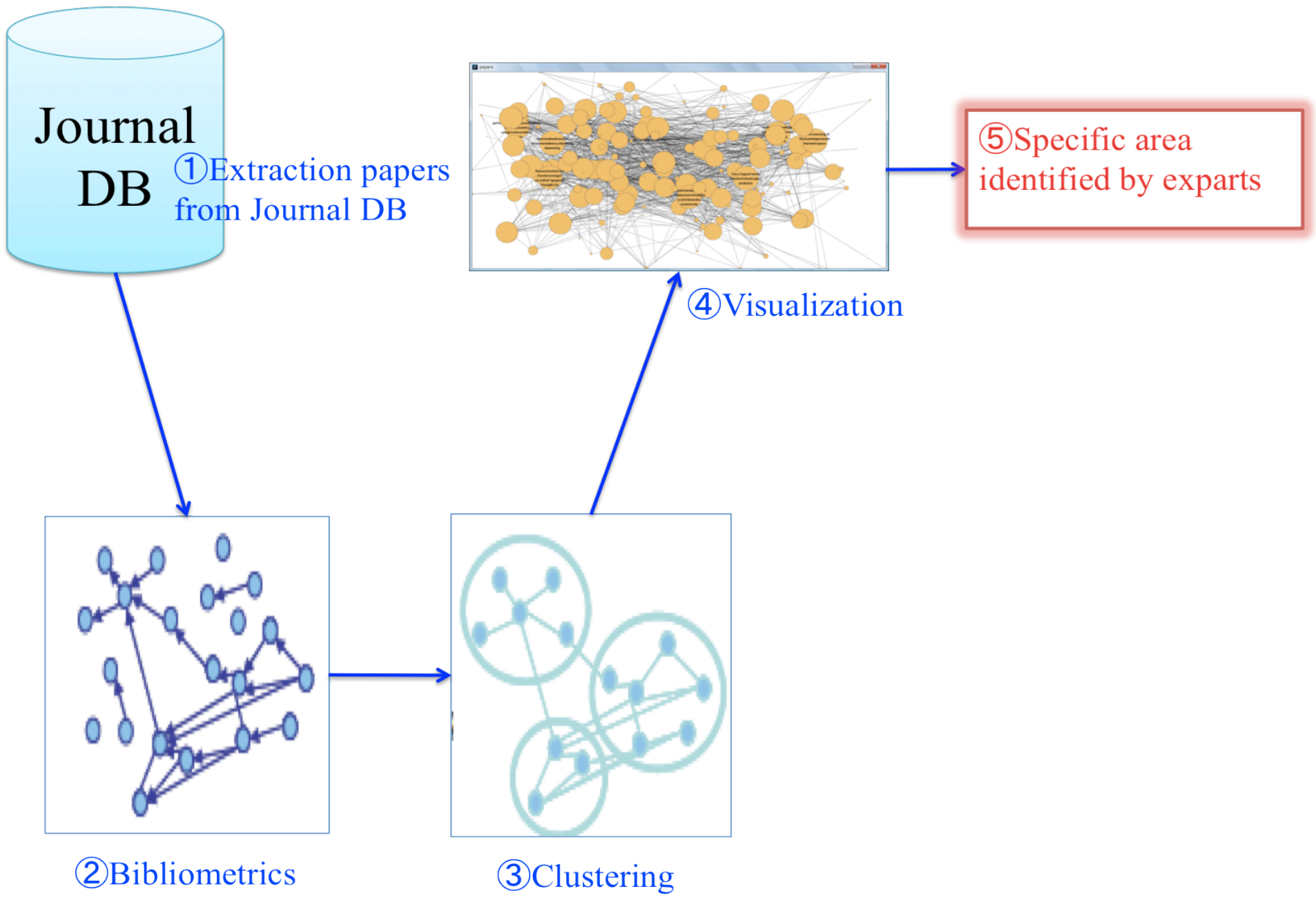

Figure 3. The Construction Process of the Academic Landscape

## 3. METHODS

To solve the problem described in the previous section, this chapter considers a method for automated extraction of key papers in each cluster. Even if the same number of papers is cited, cases such as “a large number of papers are cited within a short time” and “a small number of papers are cited over a long time” can be considered, so it is difficult to calculate the exact importance of each paper using only conventional citation analysis. In this study, therefore, we consider the cases mentioned above as directed graphs that denote papers as nodes and citations and as edges, and after assigning the publication year to each node, we attempt to calculate the importance of each paper by investigating variance values in the publication years of the original nodes of the edges that enter a certain node. The variance values will be calculated by follows formula (3).

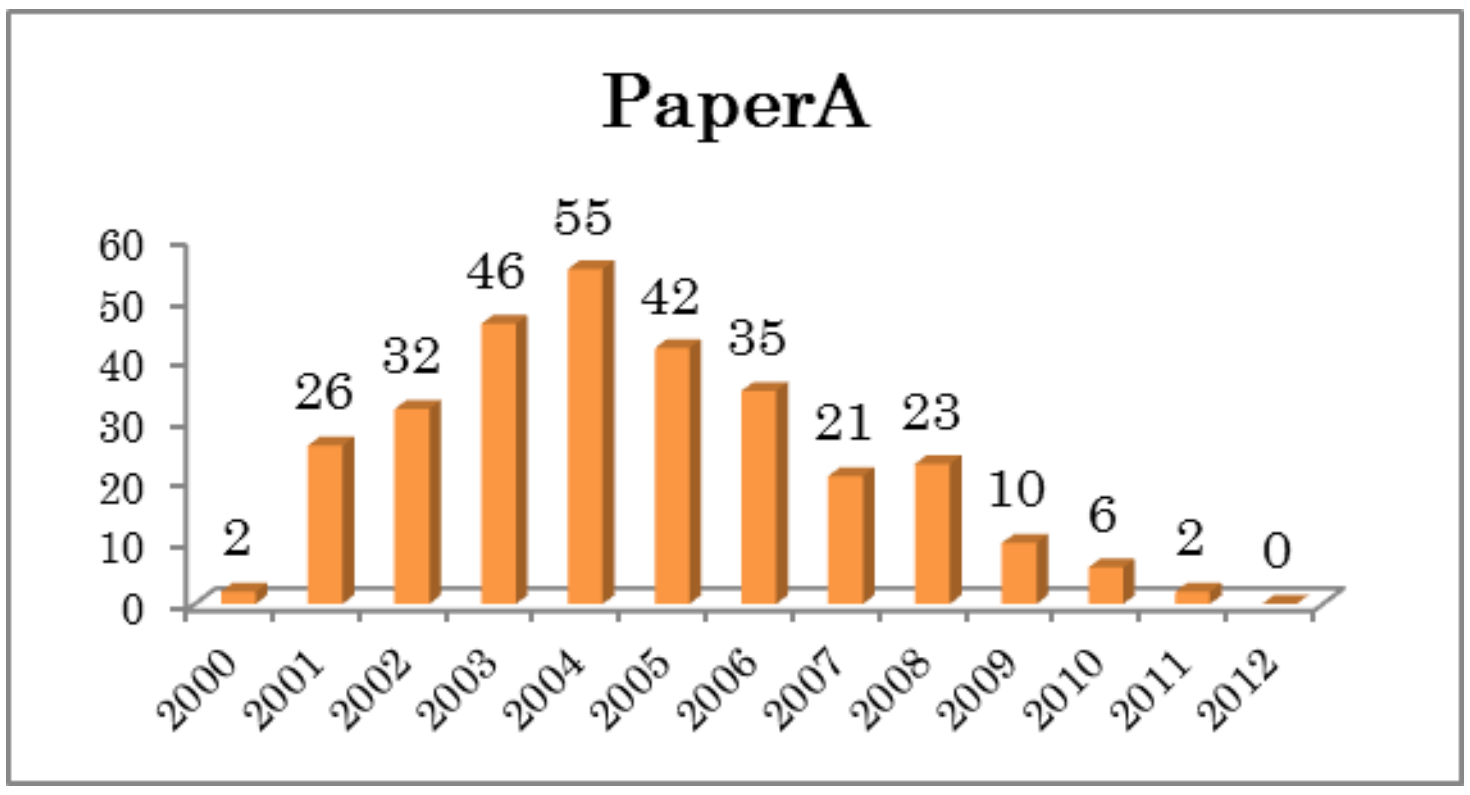

Figure 4. Example1 of transition of citation

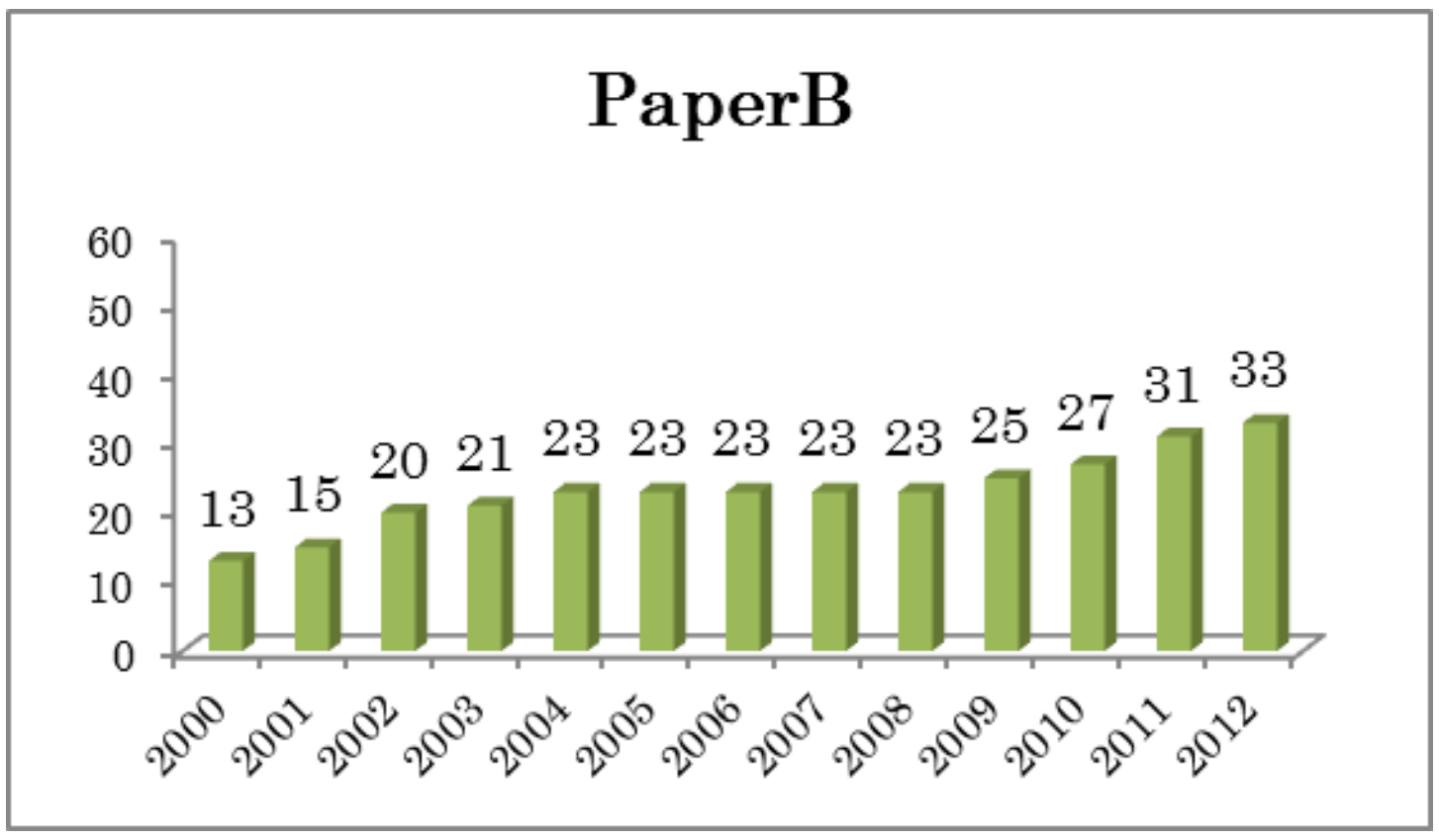

Figure 5. Example2 of transition of citation

We then attempt to create a visualized graph based on the importance. The overall flow is described below and the details of each step are described in the following sections.

1. Extract the object papers from the publication database (DB) using a keyword (query).

2. Analyze the variance values of the publication years of cited literature to give weight to each paper.

3. Apply the weight in step 2 above to the PageRank algorithm and calculate the importance of each cited paper.

4. Prepare a visualized graph based on importance.

## 3.1. Publication Databases and Search Query

In this study, Web of Science was used as the publication database. Papers were then extracted by setting "Geosciences" as the query and “1960 to 2012” as the publication years, according to the evaluation test to be described later. As a result, About 30,000 papers were extracted.

## 3.2. Publication Year Analysis of the Variance Values and Weight of Each Cited Paper

Weight was given to each paper according to the steps below.

**1) Extract the maximum value of the histogram**

Extract the year with the largest number of citations using the following function and store it as MaxYear (*MY*).

$$MY := \max\{y(x) \mid y(x) := \text{number of citations in year } y\} \quad (1)$$

**2) Specify the citation period**

Investigate years one at a time in descending order and define the year when the number of citations exceeding 10% of *MY* appeared for the first time as the starting year of the citations and store it as StartYear (*SY*). Then, store the first year when the number of citations was less than 10% of *MY* for the first time as LastYear (*LY*). Finally, obtain the period (*P*) from the start year to the last year using the following equation (2):

$$P := (LY + 1) - SY \quad (2)$$

As shown in Fig6, We will be conceivable that another trend (second trend, 2007-2010) to cite this paper was come after the first trend (1999-2005) was finished. When two or more mountains of histograms exist, repeat this operation and store the result of each operation to Period (*P*) 0, 1 to *n*.

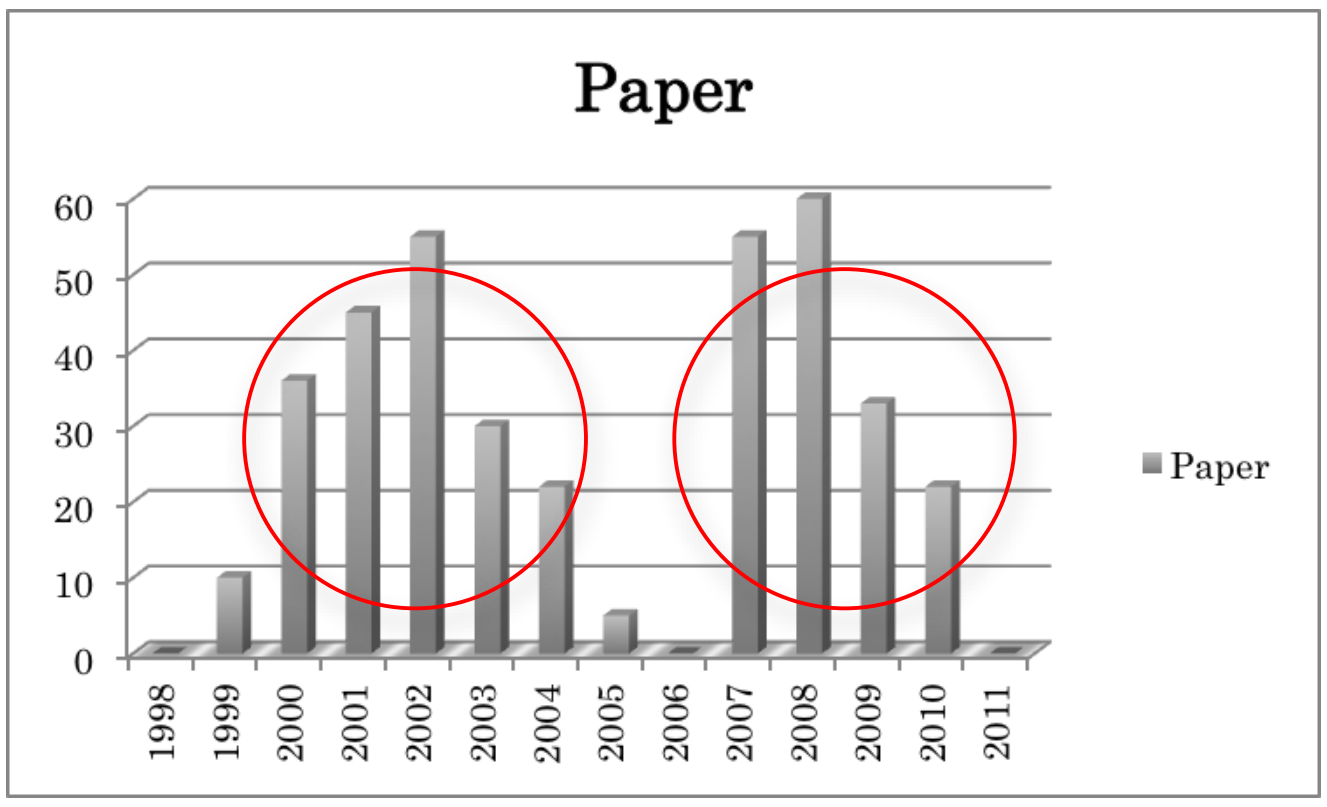

Figure 6. Example of two mountains in histograms

3) **Calculate the variance values (Standard Deviation) of the histogram**

Investigate the variance values of period in which the pertinent paper was cited by investigating the variance values (standard deviation) of the publication years of the cited papers. In this case, we method for obtaining the standard deviation is expressed as follows (3). Concretely, we assume the $P_1$, $P_2$ ・・・ $P_{n-1}$, $P_n$ as a period sample. Then, we regard a $\bar{P}$ is the arithmetic average of these. Then we will calculate variance values (3) as an arithmetic mean of $(\bar{P} - P_i)^2$.

$$S^2 = \frac{1}{n}\sum_{i=1}^{n}(\bar{P} - P_i)^2 \tag{3}$$

When we assume that $V_i$ is the paper, and $V_j$ is the paper that cite of $V_i$, and $V_k$ is the paper that cite of $V_j$., relations between $V_i$ and $V_j$ and $V_k$ and $S_i$ {(3) above} are shown in follows.

$$V_i \text{ <--} S_i\hat{}2\text{-- } V_j \text{ --} S_k\hat{}2\text{--> } V_k \tag{4}$$

"<----" and "---->" is show the direction of the edge. And formula (4) shows the variance values are attached as an attribute of the edge.

## 3.3. Dynamic Extraction of Key Paper from the Cluster Using Variance Values of Cited Literature

Page Rank Algorithm [15] determines the most important page. The Google web search engine uses this algorism. Page Rank calculates the hyperlink structures in the presence of mutual referencing relations. Page Rank integrates the impact of both inlink and outlink into one set of scores. The formula of PageRank is shown in follows.

$$PR(V_i) = (1-d) + d * \sum_{V_j \in I_n(V_i)} \frac{PR(V_j)}{|Out(V_j)|} \tag{5}$$

Where $d$ is a parameter that is set between 0 and 1. The Page Rank algorism equalizes to the total points of inlink and outlinks of each page. This total points is becomes the point of that page. It is conceivable that the page is more important in proportion to the height of the total points. We will propose the new formula ($PR^S$) by application of the above equation {above of (1)-(4)} to the PageRank.

$$PR^S(V_i) = (1-d) + d * \sum_{V_j \in In(V_i)} S_i\hat{}2 \frac{PR^S(V_j)}{\sum_{V_k \in Out(V_j)} S_k\hat{}2} \tag{6}$$

This formula incorporate into the strength of the connection between $V_i$ and $V_j$ as a weight $S_i$ and $S_k$ added to the corresponding edge that connects the two vertices. The past Page Rank algorithms equalize to the total points of inlink and outlink of each paper. Namely, the sum of the scores of the citations that inlink from each paper are equal to each other. Then, it is conceivable that the page is more important in proportion to the height of sum of the scores. But formula (6) is possible to identify the key papers in each area by applying the variance value ($S_i$) to calculation of the score of citations that incoming from each paper. Although scores have been assigned equally in the conventional algorithm when there are multiple citations that incoming the importance reflecting the state of variance values in the citation year are calculated in this study with the consideration that more citations will incoming to papers with higher variance values.

### 3.4. HAL (Highangle on Academic Landscape)

HAL (Highangle on Academic Landscape) is visualized based on the importance {formula (6)} in the previous section as shown in Fig7. For each node, the paper name is displayed and papers with higher importance assigned in the previous section are expressed by larger node sizes. Incidentally, this tool is called HAL (Highangle on Academic Landscape).

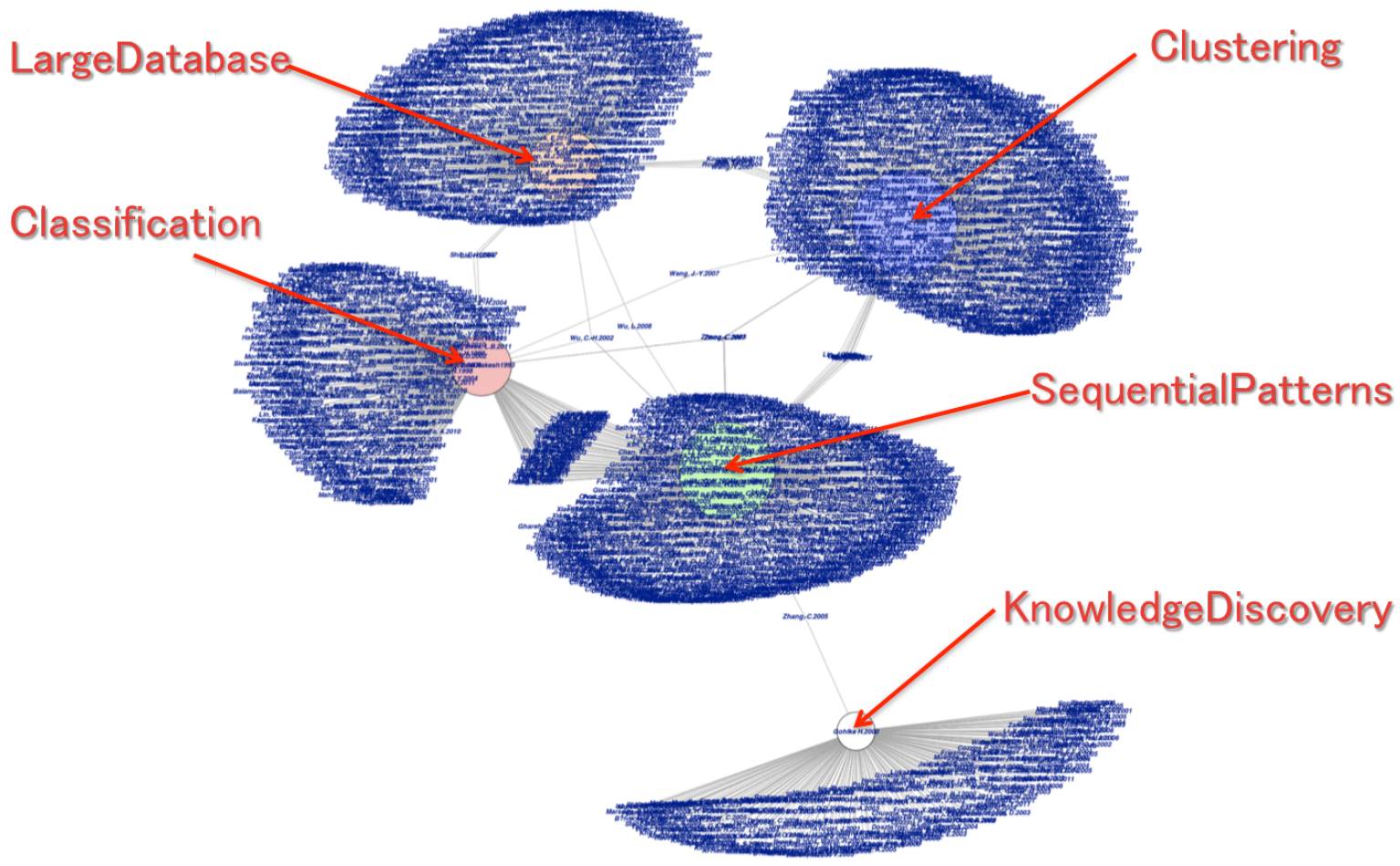

Figure 7. Example of academic landscape based on weight of each cited paper

## 4. Evaluation Experiment

### 4.1. Overview of Evaluation Experiment

This Evaluation experiment compares to this algorism (HAL) and Page Rank algorism. We show at the Table1 about details of evaluation experiment. We used the "Web of Science" as a publication database. Then we got the papers from publication database with query. Query is the "Geosciences, Multidisciplinary". It is the research area of Web of Science. We did exclude the papers that do not have a paper keywords from papers extracted. Then we were build the clustering based on the Newman method (Section2.2). The result of the clustering is Fig.5. There are 23 clusters as shown in Fig.5. In this evaluation experiment, we were used number of citations the top five papers (Table2) in the 23 clusters as a target paper of evaluation experiment.

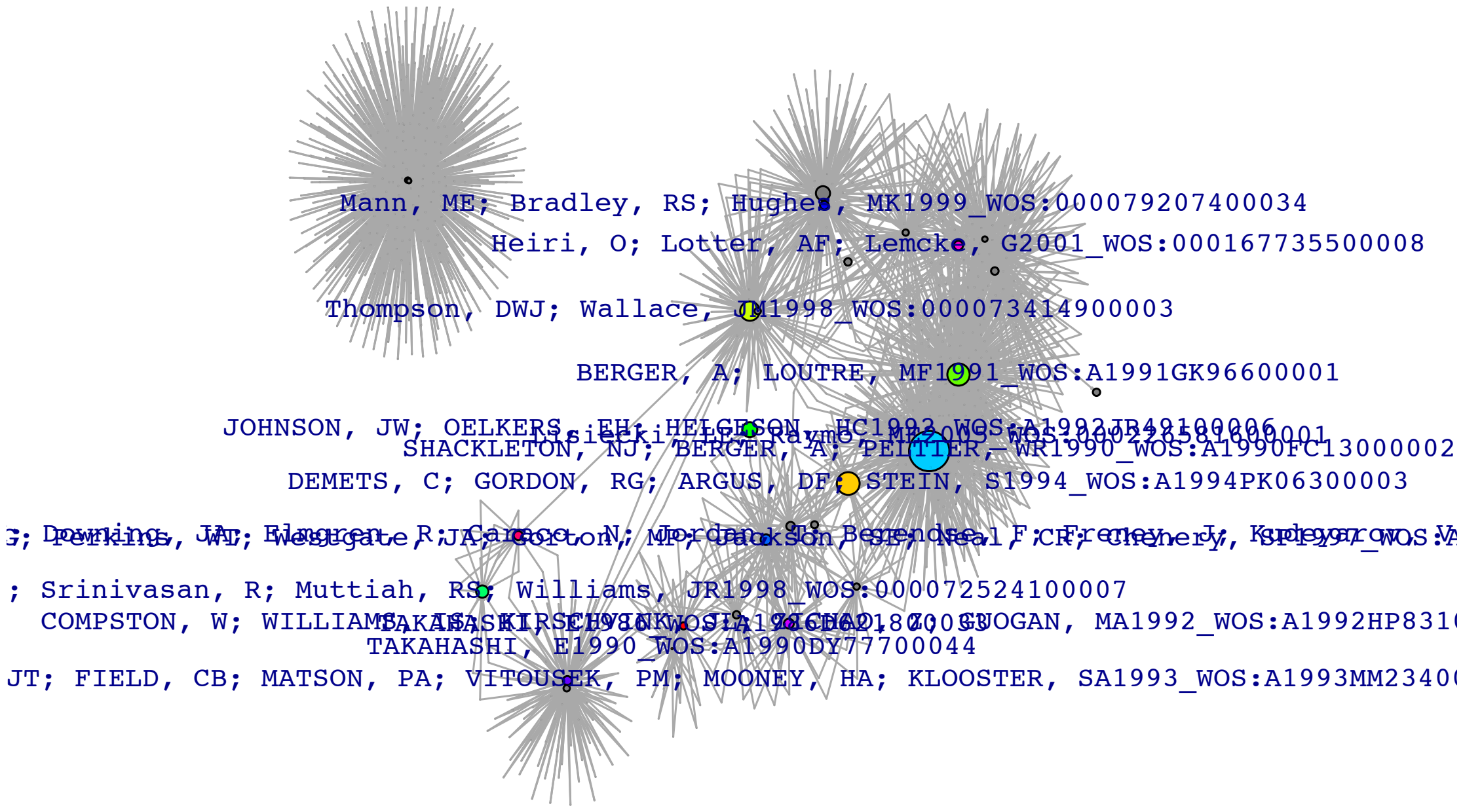

Figure.5 Result of the clustering of target papers to evaluation experiment (Query is the "Geosciences, Multidisciplinary")

Table.1 Overview of Evaluation Experiment

| | |
|---|---|
| Publication Databases | Web of Science |
| Query (Research area of Web of Science) | "Geosciences, Multidisciplinary" |
| Clustering Method | Newman method (Section2.2) |
| Calculation | R2.15 |
| Visualization | iGraph (for R2.15) |

Table.2 Number of citations the top five papers (Query is the "Geosciences, Multidisciplinary")

| Authors Name | Papers Name | Number of citations | Publication year |
|---|---|---|---|
| DEMETS, C, et.al | EFFECT OF RECENT REVISIONS TO THE GEOMAGNETIC REVERSAL TIME-SCALE ON ESTIMATES OF CURRENT PLATE MOTIONS | 1707 | 1994 |
| Thompson,DWJ, et.al | The Arctic Oscillation signature in the wintertime geopotential height and temperature fields | 1486 | 1998 |
| BERGER, A, et.al | INSOLATION VALUES FOR THE CLIMATE OF THE LAST 10000000 YEARS | 1197 | 1991 |
| JOHNSON, JW, et.al | SUPCRT92 - A SOFTWARE PACKAGE FOR CALCULATING THE STANDARD MOLAL THERMODYNAMIC PROPERTIES OF MINERALS, GASES, AQUEOUS SPECIES, AND REACTIONS FROM 1-BAR TO 5000-BAR AND 0-DEGREES-C TO 1000-DEGREES-C | 1123 | 1992 |
| Arnold, JG, et.al | Large area hydrologic modeling and assessment - Part 1: Model development | 929 | 1998 |

## 4.1. Result of Evaluation Experiment

We show the score of the result of a calculation of page rank algorithm and this study (HAL) at the Fig.6. In addition Fig.6 shows the numbers of variance values and citations of 5 papers. In this case of high variance value (Ex: "BERGER, A1991" is 29 and "JOHNSON, JW1992" is 31), the HAL's scores were larger than PageRank algorithm scores. Furthermore, although "JOHNSON, JW1992" and "Arnold, JG1998" are almost the same number of citations, the HAL's scores were different by variance values. The reason for this is the PageRank algorithm is to calculate the score from only the number of citations. Therefore the PageRank algorithm was difficult to calculate importance at following cases. Have the same number of citations, and if they are quoted at one time a large amount or has been cited to a long time little by little. But HAL is able to calculate importance these cases by considering the variance.

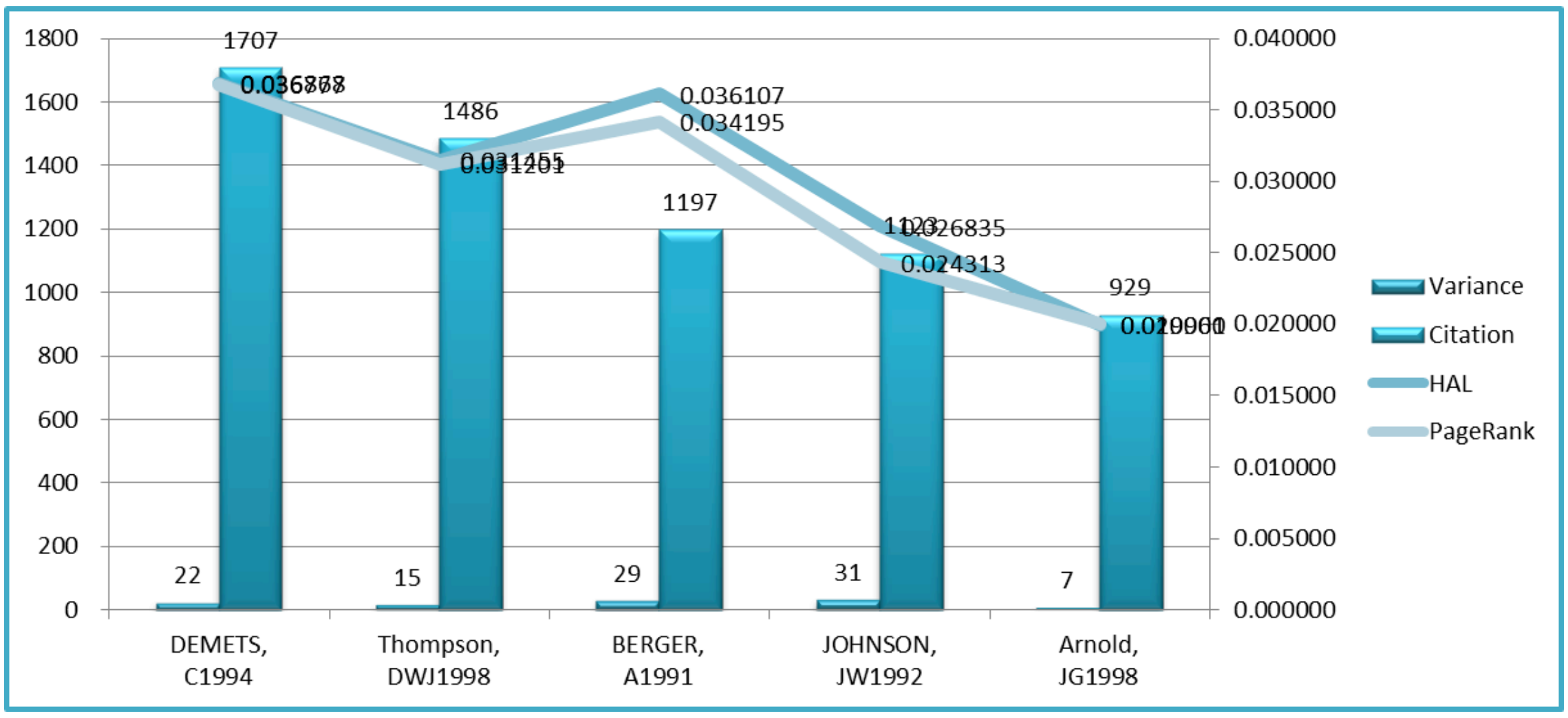

Figure.6 Compare of Page Rank algorithm and HAL (Query is the "Geosciences, Multidisciplinary")

Then, we have confirmed at another query (Geochemistry&Geophysics;Mineralogy) that this results of the evaluation experiment is not depends on the query (research area) of "Geosciences, Multidisciplinary". We selected the number of citations the top five papers (Table3) as target papers of evaluation experiment same as the evaluation experiment of "Geochemistry&Geophysics;Mineralogy". 15 clusters had been creation as shown in Fig.7. We were used numbers of citations the top five papers in the 15 clusters as a target paper of evaluation experiment. Then the result of evaluation experiment shown in Fig8. The case of high variance value ("ROEDER, PL1970" and "PEARCE, JA1979"), the HAL's scores were larger than PageRank algorithm scores. Furthermore, although "PEARCE, JA1979" and "ELLIS, DJ1979" are almost the same number of citations, the HAL's scores were different by variance values. We could confirm by this evaluation experiment that HAL was able to calculate importance these cases by considering the variance without being affected by the only number of citations.

Table.3 Number of citations the top five papers (Query is the "Geochemistry&Geophysics;Mineralogy")

| Authors Name | Papers Name | Number of citations | Publication year |
|---|---|---|---|
| KRETZ, R, et.al | SYMBOLS FOR ROCK-FORMING MINERALS | 2948 | 1983 |
| ROEDER, PL, et.al | OLIVINE-LIQUID EQUILIBRIUM | 1669 | 1970 |
| PEARCE, JA, et.al | PETROGENETIC IMPLICATIONS OF TI, ZR, Y, AND NB VARIATIONS IN VOLCANIC-ROCKS | 1389 | 1979 |
| DODSON, MH, et.al | CLOSURE TEMPERATURE IN COOLING GEOCHRONOLOGICAL AND PETROLOGICAL SYSTEMS | 1348 | 1973 |
| ELLIS, DJ, et.al | EXPERIMENTAL-STUDY OF THE EFFECT OF CA UPON GARNET-CLINOPYROXENE FE-MG EXCHANGE EQUILIBRIA | 1310 | 1979 |

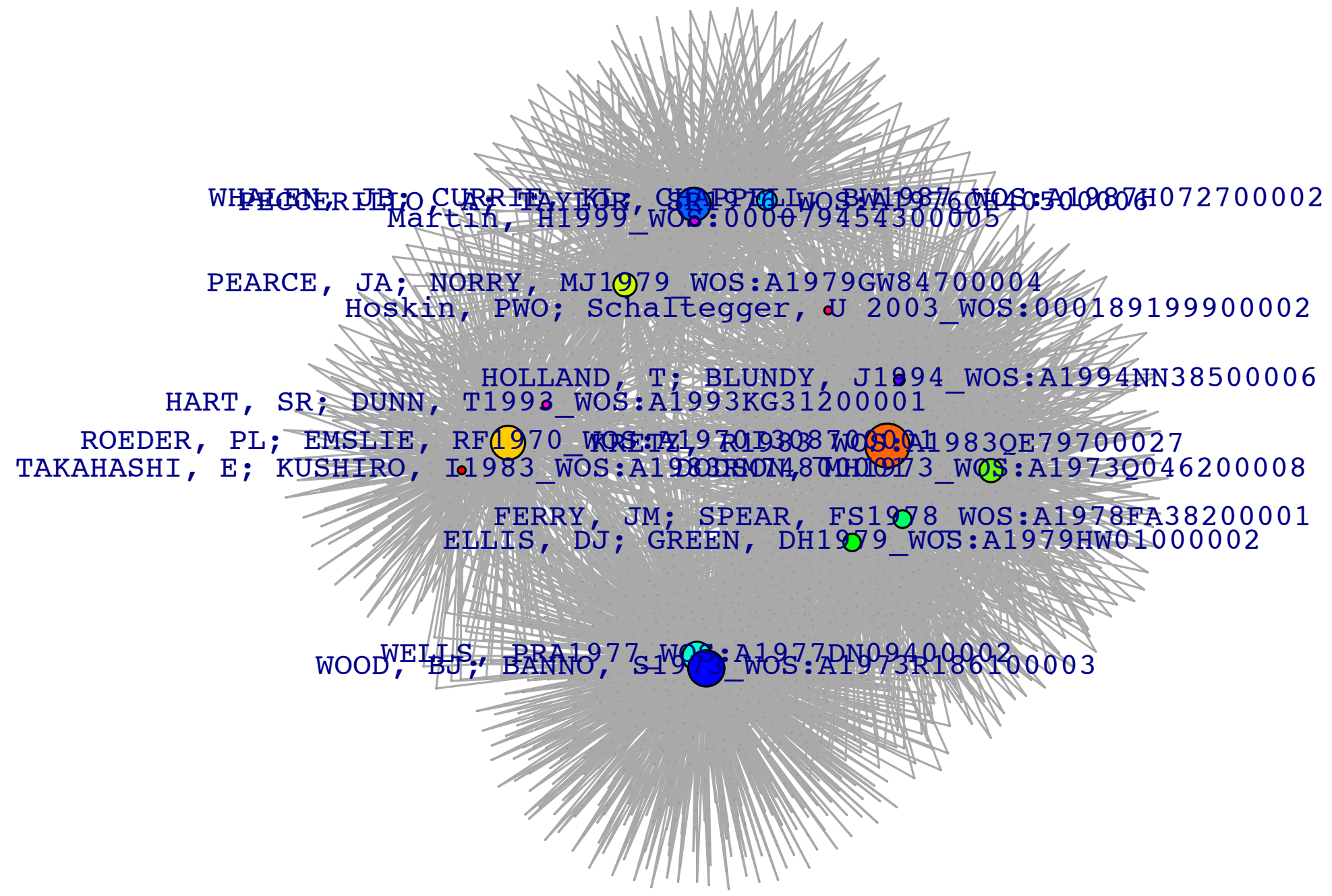

Figure.7 Result of the clustering of target papers to evaluation experiment (Query is the "Geochemistry&Geophysics;Mineralogy")

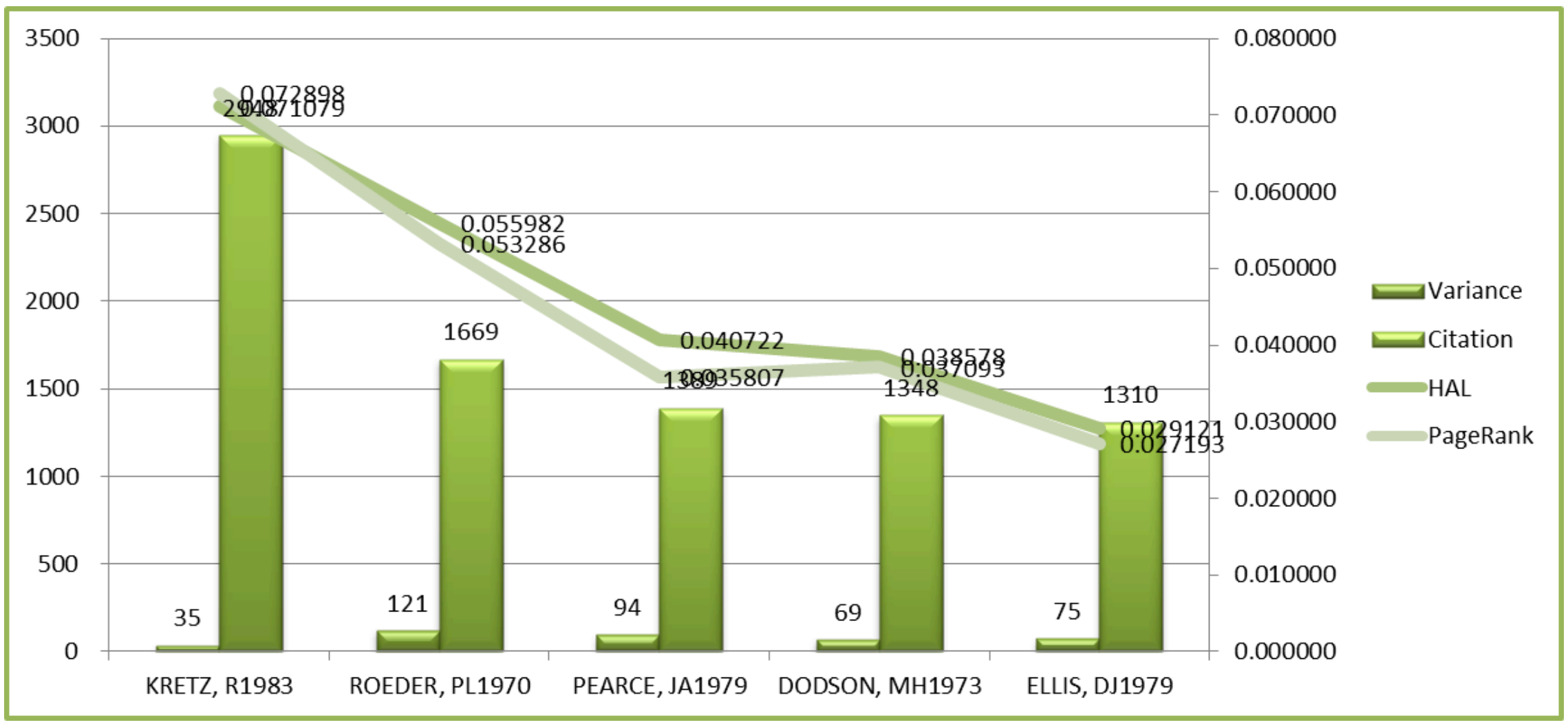

Figure.8 Compare of Page Rank algorithm and HAL (Query is the "Geochemistry&Geophysics;Mineralogy")

## 5. CONCLUSION

In this paper, we attempted to dynamically extract key papers in each area identified by clustering during analysis of citation networks of academic papers. We analyzed the variance of the publication years of the cited literature, and by applying the result to the PageRank algorithm, calculated the strictness of each paper. The PageRank algorithm was difficult to calculate importance at following cases. Have the same number of citations, and if they are quoted at one time a large amount or has been cited to a long time little by little. But this study (HAL) was able to calculate importance these cases by considering the variance. It was confirmed by this evaluation experiment.

## REFERENCES

[1] R・J・Wilson (2001). Introduction to Graph Theory, Kindai Kagaku Sha Co.,Ltd.

[2] Burt, Ronald S. (1992). Structural Holes, The Structure of Competition. Cambridge, MA: Harvard University Press.

[3] Sabidussi, G. (1966). The centrality index of a graph, Psychometrika 31, pp.581-603.

[4] Opsahl, Tore; Agneessens, Filip; Skvoretz, John (2010). Node centrality in weighted networks, Gen-eralizing degree and shortest paths, Social Networks 32: 245.

[5] Amaral, L.A.N. et al (2000). Classes of small-world networks, Proceedings of the National Academy of Sciences of the United States of America 97, No. 21, pp.11149-11152.

[6] Duncan J. Watts (2003). Small Worlds, The Dynamics of Networks Between Order and Randomness, Princeton Studies in Complexity.

[7] M. E. J. Newman and M. Girvan (2004). Finding and evaluating community structure in networks, Physical Review E, Vol. 69.

[8] M. E. J. Newman (2004). Fast algorithm for detecting community structure in networks, Phys. Rev. E, Vol. 69.

[9] M. E. J. Newman (2005). A measure of betweenness centrality based on random walks, Social Networks, Vol. 27, No.1, pp.39-54.

[10] Garfield, E. (1955). Citation Indexes for Science, Science 122, pp.108-111.

[11] Garfield, E., Welljams-Dorof, A. (1992) Of Nobel class: A citation perspective on high impact research authors, Theoretical Medicine 13 (2), pp.117-135.

[12] Garfield, E. (2000). Use of Journal Citation Reports and Journal Performance Indicators in measuring short and long term journal impact, Croatian Medical Journal 41 (4), pp.368-374.

[13] H. Small (1973). Co-citation in the scientific literature: a new measure on the relationship between two documents, Journal of the American Society for Information Science, Vol. 24, pp.28-31.

[14] M. Kessler (1963). Bibliographic coupling between scientific papers, American Documentation Volume 14, Issue 1, pp.10–25.

[15] S. Brin and L. Page. (1998). The anatomy of a large-scale hypertextual Web search engine. Computer Networks and ISDN Systems, 30(1–7).

**Authors**

### Akira Otsuki

Received his Ph.D. in engineering from Keio University (Japan), in 2012. He is currently associate professor at Tokyo institute of technology (Japan) and Officer at Japan society of Information and knowledge (JSIK). His research interests include Analysis of Big Data, Data Mining, Academic Landscape, and new knowledge creation support system. Received his Best paper award 2012 at JSIK. And received his award in Editage Inspired Researcher Grant, in 2012.

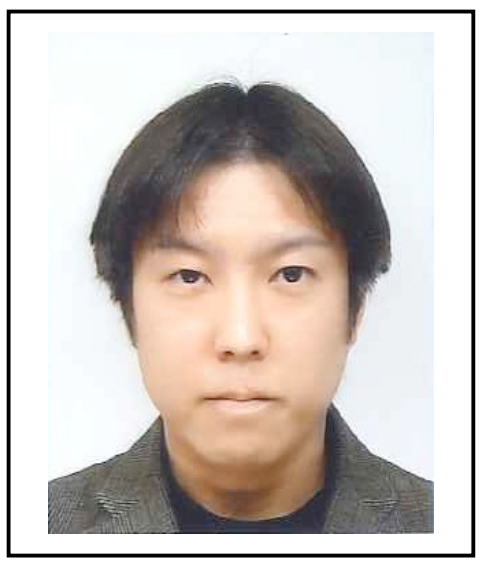